\documentclass[aps,prd,epsf,twocolumn,showpacs,groupedaddress,nofootinbib]{revtex4}
\usepackage{graphicx}
\usepackage{amsmath,amssymb,latexsym}
\usepackage{bm}

\newcommand{\gsim}{\mathrel{\vcenter{\hbox{$>$}\nointerlineskip\hbox{$\sim$}}}}

\newcommand{\nutau}{\nu_\tau}
\newcommand{\nue}{\nu_e}
\newcommand{\numu}{\nu_\mu}
\newcommand{\Enue}{E_{\nu_e}}
\newcommand{\Enumu}{E_{\nu_\mu}}

\newcommand{\Enuj}{E_{\nu_j}}
\newcommand{\Fnue}{F_{\nu_e}}
\newcommand{\Fnumu}{F_{\nu_\mu}}

\newcommand{\Fnuj}{F_{\nu_j}}
\newcommand{\Epho}{E_\gamma}
\newcommand{\Epi}{E_\pi}

\newcommand{\Fpho}{F_\gamma}
\newcommand{\bwide}{\begin{widetext}}
\newcommand{\ewide}{\end{widetext}}
\newcommand{\beq}[1]{\begin{equation} \label{(#1)}}
\newcommand{\eeq}{\end{equation}}
\newcommand{\ba}[1]{\begin{eqnarray} \label{(#1)}}
\newcommand{\ea}{\end{eqnarray}}

\begin{document}
\title
{Neutrino Bursts from Fanaroff-Riley I Radio Galaxies}
\author{Luis A. Anchordoqui$^a$, Haim Goldberg$^a$, Francis Halzen$^b$, and
Thomas J. Weiler$^{c,d}$}
\affiliation{$^a$Department of Physics, Northeastern University, Boston MA 02155\\
$^b$Department of Physics, University of Wisconsin, Madison WI 53706\\
$^c$Theory Division, CERN, CH-1211 Geneva 23, Switzerland\\
$^d$Department of Physics and Astronomy, Vanderbilt University, Nashville TN 37235
}

\date{\today}
\begin{abstract}
On the basis of existing observations (at the 4.5 $\sigma$ level)
of TeV $\gamma$-ray outbursts from the Fanaroff-Riley I (FRI)
radio galaxy Centaurus A, we estimate the accompanying neutrino
flux in a scenario where both photons and neutrinos emerge from
pion decay. We find a neutrino flux on Earth $dF_{\nu}/dE_\nu =
4.5\times 10^{-11}\ (E_\nu/{\rm TeV})^{-2}\ {\rm TeV}^{-1}\ {\rm
cm}^{-2}\ {\rm s}^{-1},$ equally spread in flavor as a result of
maximal mixing. Such a flux will trigger at the IceCube facility
about 10 showers/burst, with negligible background from
atmospheric muons, and  primary neutrino energies in excess of
100 TeV. The only other FRI radio galaxy observed in the TeV
photon energy range at the 4$\sigma$ level is M87. The burst
nature of this activity is not established; however, we show that
the intrinsic neutrino luminosity during the active period is the
same as the Centaurus A burst. On the assumption that Centaurus A
typifies the FRI population, we show that IceCube should collect
10 showers (all neutrino flavors) in 3 years, attaining a  95\%~CL
sensitivity to the diffuse neutrino flux from FRI radio galaxies
in one year of observation.
\end{abstract} 

\maketitle

Conventional astronomy spans about 18 decades in photon
wavelengths, from $10^4$~cm radio-waves to $10^{-14}$~cm
$\gamma$-rays of GeV energy. Because the universe is opaque to photons of
TeV energy and above, present studies focus on hadrons,
neutrinos, and gravitational waves as messengers probing the high
energy universe. The best candidates to serve as messengers in a
new astronomy of the high energy behavior of distant sources are
neutral particles. This is because the orbit of a charged cosmic
ray can be substantially bent by the ambient magnetic field of
our own galaxy, destroying the possibility of locating the source.
The most promising messenger is the neutrino: it can be copiously
produced in cosmic beam dumps and can traverse unscathed dense
astrophysical environments. In this context, we recently carried
out a study of directional neutrino signals from sources in the
Galactic plane~\cite{Anchordoqui:2003vc}. In this letter we extend this project
to the extragalactic domain, and delineate the prospects to
identify Fanaroff-Riley I (FRI) radio galaxies as high energy neutrino emitters.
In our discussion it will be
seen explicitly how the neutrinos provide a signal precisely
where photon signals are blocked by interactions with the universal radiation
backgrounds permeating the universe.

At present, a handful of sources have been established as TeV
$\gamma$-ray emitters~\cite{Punch:xw}. All of them are nearby BL
Lac objects characterized by strong rapid variability and
apparent superluminal motion. These extreme features are
generally interpreted as a consequence of dissipative effects
(non-thermal emission) from a relativistic jet oriented at small
angle with respect to the line-of-sight~\cite{Blandford}.

There are two principal mechanisms for TeV gamma ray production:
{\it (i)}~Electrons  undergo bremsstrahlung in the
magnetic field and/or inverse Compton scattering in the ambient
photon sea or {\it (ii)}~the gamma rays are directly traced to $\pi^0$
decay. Only the second scenario can accommodate baryonic cosmic ray
production. Since such cosmic rays are observed, it is reasonable to assume that
at least some  gamma ray sources operate according to the second mechanism.

In the context of unification models~\cite{Urry:1995mg}, BL Lac
objects are intrinsically the same as FRI. Though the jet
emission from FRIs is not strongly Doppler boosted towards us
(and may even be de-boosted), in some cases the lack of
relativistic boosting can be partially compensated by
proximity to Earth. Therefore, some FRI radio galaxies could be
detectable in the TeV range. In what follows, we examine the consequences
of assuming that gamma ray emission from FRI originates in $\pi^0$ decay,
and is necessarily accompanied by a flux of high energy neutrinos emerging
from the $\pi^{\pm}$ population.

At a distance of 3.4 Mpc, Centaurus A (Cen A) is the
prototype of FRI galaxies~\cite{Israel:1998ws}. It is the only GeV
gamma ray source with a confirmed large-inclination
jet~\cite{Sreekumar:1999xw}. Data collected in the early 70's
with the optical intensity interferometer operated by Sydney University
at Narrabri, show a $\sim 4.5\sigma$ (time average) excess of $\gamma$-ray
events from the direction of Cen A~\cite{Grindlay}.
The reported cumulative flux averaged over 3 yr of observations is
\begin{equation}
F_\gamma (E_\gamma > {\rm 300}~{\rm GeV}) = 4.4 \pm 1.0 \times 10^{-11}\,\,
{\rm cm}^{-2}\,\, {\rm s}^{-1}\,.
\label{fgamma}
\end{equation}
Based on a power-law spectrum $\propto E_\gamma^{-2},$ this
corresponds to a luminosity at the source of ${\cal L}\approx 7.5
\times 10^{40}$ erg s$^{-1}$. It is important to note that since
the detector beam did not include the radio lobes, it is likely
that the source resides in the compact nucleus of Cen A. The data
reveals two different periods of heightened activity in the TeV
range, each of about 1 year duration. Though data at the upper
end of the spectrum await confirmation, the $\gamma$-ray
bandwidth of Cen A could reach energies as high as hundreds of
TeV. There are specific hints in this direction in data reported
during the 1980's and 1990's by the Buckland~\cite{Clay:uy} and
JANZOS~\cite{Allen:1993ir} collaborations. Several items can be
noted with respect to these measurements: {\it (i)} The data are
consistent with bursts of approximately one year duration every
decade. {\it (ii)} Both experiments reveal an upper cutoff of
$\sim 200$ TeV, which can be ascribed to absorption on the
background radiation fields. {\it (iii)} The reported flux in both
cases is about 2 orders of magnitude larger than that measured at
the Narrabri Observatory; the chance probabilities are $6\times
10^{-3}$ and $2\times 10^{-2}$ for Buckland and JANZOS,
respectively. Although these data are not at the discovery level,
the observations of 100-TeV photons at the two facilities
corroborate one another. {\it (iv)} The emission of 100-TeV gamma
rays is supportive of their origin in pion production and decay
processes, as opposed to synchrotron and inverse Compton
scattering~\cite{Anchordoqui:2001nt}.

Additional relevant information about the source is contained in
these observations. The $\sim$ 1-year duration of the observed burst at
Narrabri implies a coherent region of activity $r \sim 0.3$~pc. Therefore,
the recent discovery of a sub-parsec radio counterjet in the nucleus of
Cen A is of interest~\cite{Jones:1996si}.
As a conservative estimate we adopt the flux in Eq.(\ref{fgamma})
as the  photon excess characterizing the Cen A burst, which
corresponds to an integrated burst energy
$E_{\rm burst}= 2.3\times 10^{48}$~erg.

A high-energy neutrino flux emerges if the charged pion decay
length is smaller than the pion interaction length in the source
region. The latter, as well as the pion production rate, depends
on the gaseous hadronic particle density. An estimate of the
target nucleons can be obtained by assuming that the gas density
$n$ near the accretion disk of the supermassive black hole engine
of Cen A is similar to that near the center of our Galaxy. This
in turn is arrived at by assuming that the gas/star density ratio
near the center is the same as the one in the Galactic disk,
yielding $n \approx 10^6$~cm$^{-3}$. Our estimate  is most likely
a lower bound on the gas density, since the Cen A black hole
mass  $M_{\rm BH} \approx 2\times 10^8\
M_{\odot}$~\cite{Marconi:2000tz} is about 2 orders of magnitude
greater than SgrA$^*$~\cite{Ghez:1998ph,note1}. With this in mind, the
corresponding pion mean free path $(n\sigma_{\pi N})^{-1} \approx
6.5$~pc is much greater than the charged pion decay length, which
at 100~TeV is $5.6\times 10^8$~cm. (Here $\sigma_{\pi N} \approx
50$~mb is the pion-nucleon cross section.) On the other hand, the
mean free path for collision of the ultrarelativistic accelerated
protons on the gas  is $(n\sigma_{pN})^{-1} \approx 4.6\ {\rm
pc}$, where $\sigma_{p N} \approx 70$~mb~\cite{note2}. Assuming no
significant
deflections on the magnetic field, this implies a probability of
interaction in the coherent region $p = r\ n \ \sigma_{pN}\approx
7\%.$ The infalling mass $M_{\rm infall}$ required to power the
burst is found through the relation
\begin{equation}
E_{\rm burst} = p \,G\,M_{\rm BH} \,M_{\rm infall}/r_s\,\,,
\end{equation}
where $G$ is Newton's constant and $r_s$ is the Schwarzschild radius.
This gives $M_{\rm infall}\approx 3.6\times 10^{-5}\ M_{\odot}$.
This is about 100 times larger than the estimated infall
rate for the Galactic black hole~\cite{Genzel:2003cn}.
Because of the burst nature, coupled with the
much larger mass for the Cen A black hole, we take this as a reasonable
accretion rate, allowing sufficient pion production.

Since $\pi^0$'s, $\pi^+$'s, and $\pi^-$'s are made in equal
numbers, one expects two photons, two $\nue$'s, and four
$\numu$'s per $\pi^0$. On average, the photons carry one-half of
the energy of the pion, and the neutrinos carry one-quarter. The
energy-bins $dE$ scale with these fractions, and we arrive at
\begin{eqnarray}
\frac{d\Fpho}{d\Epho} (\Epho=\Epi/2) & = &
    4\,\frac{dF_{\pi}}{d\Epi}(\Epi)\,, \nonumber \\
\frac{d\Fnue}{d\Enue} (\Enue= \Epi/4 ) & = &
    4\,\frac{dF_{\pi}}{d\Epi}(\Epi)\,, \\
\frac{d\Fnumu}{d\Enumu} (\Enumu= \Epi/4 ) & = &
    8\,\frac{dF_{\pi}}{d\Epi}(\Epi)\,,\nonumber
\end{eqnarray}
for the fluxes at the source, where $\pi$ denotes any one of the
three pion charge-states. Terrestrial experiments (see
e.g.~\cite{Fukuda:1998mi}) have shown that $\numu$ and $\nutau$
are maximally mixed with a mass-squared difference $\sim
10^{-3}{\rm eV}^2$. This together with the known smallness
of $|\langle \nu_e|\nu_3\rangle|^2$,
implies that the $\numu$'s will partition themselves equally
between $\numu$'s and $\nutau$'s on lengths
large compared to the oscillation length $\lambda_{\rm osc}\sim
1.5 \times 10^{-3}\,({E_\nu}/{\rm PeV})$~pc.
Here $\nu_3 \simeq (\nu_\mu+\nu_\tau)/\sqrt{2}$
is the third neutrino eigenstate. From
these remarks,
one finds a nearly identical flux for each of the three neutrino
flavors ($j=e,\mu,\tau$), which is equal to
\begin{equation}
\frac{d\Fnuj}{d\Enuj} (\Enuj= \Epho/2) = 2\frac{d\Fpho}{d\Epho}(\Epho)\,.
\label{fnu}
\end{equation}
On the assumption of an $E_{\gamma}^{-2}$ spectrum, we fix the
normalization using the cumulative number flux of Eq.~(\ref{fgamma}) with
an upper cutoff at 3 TeV~\cite{Grindlay}. From Eq.~(\ref{fnu}) we then
obtain the neutrino flux on Earth
\begin{equation}
\frac{dF_{\nu_j}}{dE_\nu} = 1.5
\times 10^{-11} \,\, \left( \frac{E_\nu}{{\rm TeV}} \right)^{-2} \,\, {\rm
TeV}^{-1} {\rm cm}^{-2} {\rm s}^{-1} \,.
\label{fnumu}
\end{equation}
The observation of 100-TeV photons supports the extrapolation of
the spectrum in Eq.~(\ref{fnumu}) to those high energies.

The most promising prospect for detection of such a low flux is the
IceCube facility under construction at the South Pole~\cite{Ahrens:wv}.
For energies above $\sim 100$~TeV,  IceCube will
provide $4\pi$ detection of neutrinos since the atmospheric muon
flux is negligible. Moreover, at these energies the telescope
will be able to resolve
directionality in all neutrino flavors. The total number of showers
in a single burst of duration $t\simeq$ 1 yr  is then
\begin{equation}
{\cal N}_{\rm IceCube} = N_A \,\,\, \rho \,\,t\,\, V_{\rm eff} \,
\sum_j \int_{E_{\nu, {\rm min}}}^ \infty \frac{dF_{\nu_j}}{dE_\nu}\
\sigma^{{\rm CC}}_{\nu N}\ dE_{\nu}\  ,
\label{cenaice}
\end{equation}
where $N_A$ is Avogadro's number,
$V_{\rm eff} \approx 2$~km$^3$ is the effective volume of ice with density
$\rho,$ and $\sigma^{{\rm CC}}_{\nu N}=6.78\times 10^{-35}\
(E_{\nu}/{\rm TeV})^{0.363}$~cm$^2$ is the charged
current  neutrino-nucleon cross section~\cite{Gandhi:1998ri}. By substituting
the flux given in Eq.(\ref{fnumu}), we find about 10 showers per burst,
with primary neutrino energy $> 100$~TeV. Thus we arrive at the first prediction of this paper:
{\em during a burst similar to the 4.5$\sigma$ event detected by the Narrabri
Observatory, IceCube will detect about one 100-TeV neutrino per month pointing
to Cen A. This signal persists even in the absence of the emitted photons
which are absorbed on the infrared and microwave radiation backgrounds.}

Located in the Northern hemisphere at a distance of 16~Mpc, M87
is the FRI with the brightest optical jet~\cite{Roser}. It shows
most of the characteristics of BL Lac objects, with the jet
oriented at $30^\circ - 35^\circ$ to our line of sight. Data
taken during 1998-1999 with the HEGRA stereoscopic system of 5
imaging atmospheric \v{C}erenkov telescopes show an excess of
photons from the direction of M87 with significance level $4.1\sigma$ above
background~\cite{Aharonian:2003tr}. The observed
cumulative  flux is
\begin{equation}
F_\gamma (E_\gamma >  730~{\rm GeV}) = 0.96 \pm 0.23 \times 10^{-12}\,\,
{\rm cm}^{-2}\,\, {\rm s}^{-1}\, .
\label{hegra}
\end{equation}
The data can be fit with a power law $dF_{\gamma}/dE_{\gamma}\sim
E_{\gamma}^{-\alpha},$ with $\alpha = 2.9\pm 0.8_{\rm stat}\pm
0.08_{\rm syst}.$ The large uncertainty in the spectral index is
not, however, reflected in obtaining the source luminosity: with
an upper cutoff of 5 TeV, there is about a 25\% variation as
$\alpha$ varies between 2 and 2.9. Hereafter, we take $\alpha =2$
to reflect the conventional Fermi engine emission spectrum. This
corresponds to a $\gamma$-ray luminosity at the source ${\cal L}
\approx 6.8\times 10^{40}$~erg s$^{-1}.$ This is remarkably close
to the $\gamma$-ray source luminosity of the Cen A burst obtained
above.

By duplicating for M87 our Cen A
discussion, we obtain
\begin{equation}
\frac{dF_{\nu_j}}{dE_\nu} =  7 \times 10^{-13} \,\,
\left( \frac{E_\nu}{{\rm TeV}} \right)^{-2} \,\, {\rm
TeV}^{-1} {\rm cm}^{-2} {\rm s}^{-1} \,.
\end{equation}
A rough estimate of the event rate at IceCube can be obtained following
the analysis of our previous paper~\cite{Anchordoqui:2003vc}. Using the
terrestrial transmission probability given in~\cite{Gaisser:1994yf}, and
a conservative effective area of 1 km$^2$,
we find about 2 events/yr during a period of TeV gamma ray activity. This is
similar to the $\nu_{\mu}$ atmospheric background. A more definitive
assessment of sensitivity
to such a signal will await
further refinement of angular and energy resolutions via improved knowledge of
the detector response. Additionally, the temporal profile of future gamma ray
observation will allow a better delineation of the signal.

Although there are no other nearby  FRIs of this
magnitude which can potentially be detected as point sources, one
can integrate over the estimated FRI population out to the
horizon to obtain a prediction for the diffuse neutrino flux.
This quantity is given by
\begin{equation} J_\nu (E_\nu) = \frac{1}{4\pi} \,\,{\cal R}\,
\,n_{_{\rm FRI}} \,\, {\cal L}_{\nu} \label{j}
\end{equation}
where ${\cal R} \simeq$~1 horizon $\simeq 3$~Gpc, $n_{_{\rm FRI}}
\sim 8 \times 10^4$~Gpc$^{-3}$ is the number density of
FRI~\cite{Padovani}, and ${\cal L}_\nu = (dN_0/dE_{\nu})/\tau$ is
an average  neutrino luminosity (all flavors) of FRI radio
galaxies. Here $dN_0/dE_{\nu}$ is the differential injection
spectrum for a single burst and $\tau
\sim 10$~yr is the period between outbursts. Although the burst nature of the HEGRA
observation is not established, in what follows we conservatively assume that the
photon excess from M87 is localized to an outburst period.
Since both Cen A and M87 have almost
the same luminosity, we adopt this as the average for FRI radio galaxies.
Thus, the quantity
$dN_0/dE_{\nu}$ can be obtained from Eq.~(\ref{fnumu}),
\begin{equation}
\frac{dN_0}{dE_{\nu}} = 4\pi d^2\ N_f \ t\ \frac{d\Fnuj}{dE_\nu}\,
\label{dnde}
\end{equation}
where $N_f=3$ is the number of neutrino flavors and $d$ is the
distance to the source. For long time averages we find
\begin{equation}
\frac{1}{\tau}\ \frac{dN_0}{dE_{\nu}} = 5.7 \times 10^{39}\, E_\nu^{-2}\,\,\,\, {\rm TeV}^{-1}\,
{\rm s}^{-1}\ . \label{dndea}
\end{equation}
The diffuse flux observed on Earth then follows from
Eq.~(\ref{j})
\begin{equation}
E_{\nu}^2\ J_\nu \simeq
1.2 \times 10^{-11}\, {\rm TeV} \ {\rm cm}^{-2} \ {\rm s}^{-1} \ {\rm sr} ^{-1}\,.
\end{equation}
Note that this flux is about a factor of 2 smaller than the
Waxman-Bahcall~\cite{Waxman:1998yy} upper limit on the intensity
of neutrinos produced in sources which also emit baryonic cosmic
rays. After 3 years of observation, the 90\% CL sensitivity
(corresponding to 2.44 events) of IceCube to the diffuse muon
neutrino flux is $E_{\nu}^2\ J_{\nu_{\mu}}\simeq 3\times
10^{-12}\, {\rm TeV} \ {\rm cm}^{-2} \ {\rm s}^{-1} \ {\rm sr}
^{-1}\,$~\cite{Ahrens:wv}. We arrive then at the second
prediction of this work: {\em If the observations of Cen A at
Narrabri and  HEGRA  measurements of M87 characterize the
outburst of FRI radio galaxies, then  IceCube will collect about
10 neutrino events (all flavors) in three years.}

It seems worthwhile to briefly examine the implications of
neutrino bursts of magnitude corresponding to the $\gamma$-ray
intensities reported by the Buckland and JANZOS collaborations,
with full knowledge that these observations are not at the
discovery level. Such a powerful burst implies a neutrino flux on
Earth about 2 orders of magnitude greater than that in
Eq.~(\ref{fnumu}). In addition to vastly increasing the event
rate at IceCube, the $\nu_\mu$ flux should be detected by 0.1
km$^2$ neutrino telescopes under construction
(ANTARES~\cite{Aslanides:1999vq}, NESTOR~\cite{Tzamarias:wd}) and
a planned 1 km$^3$ facility~\cite{NEMO}, all in the Mediterranean.
The location of ANTARES at 43$^\circ$ North provides a 5$\sigma$
discovery sensitivity of $1.2\times 10^{-10}\ (E_{\nu}/{\rm
TeV})^{-2}$ TeV$^{-1}$ cm$^{-2}$ s$^{-1}$ (for muon neutrinos) in
the direction of Cen A~\cite{Blanc:2003mu}. We find that in the
course of $\sim $ 10 years of observation, ANTARES and NESTOR
will measure a neutrino burst an order of magnitude larger than
the 5$\sigma$ detector sensitivity.

In summary, we have analyzed the possibility of detecting the
neutrino counterparts of various TeV $\gamma$-ray observations in
a model where  gamma rays originate through $\pi^0$-decay at the
source. We have found that Icecube will attain sensitivity to
observe neutrino bursts from Cen A. Moreover, if the $\gamma$-ray
observations with the Narrabri Observatory and the HEGRA \v{C}erenkov telescopes
characterize the emission behavior of the FRI population, then IceCube should
observe the diffuse $\nu_{\mu}$ flux with
statistical significance in several years. In light of this, we
conclude that observations at future neutrino telescopes will permit a major advance in
discriminating between high energy astrophysical processes.

We would like to thank Felix Aharonian for valuable discussions.
This work has been partially supported by US NSF (Grant Nos.
OPP--0236449, PHY--0140407, PHY--0244507), US DoE (Grant Nos.
DE-FG02-95ER40896, DE-FG05-85ER40226), NASA-ATP 02-000-0151, the
Wisconsin Alumni Research Foundation, and Vanderbilt University's
Discovery Grant and sabbatical programs.


\begin{thebibliography}{99}
\bibitem{note2} The
interaction of hadrons with the thermal photon background at
$T \alt 10^4$ K is completely negligible.
\end{thebibliography}
\end{document}